\documentstyle[aps,epsf,psfig,prb,preprint,tighten]{revtex}

\newcommand{\be}{\begin{equation}}
\newcommand{\ee}{\end{equation}}
\newcommand{\bc}{\begin{center}}
\newcommand{\ec}{\end{center}}
\newcommand{\bea}{\begin{eqnarray}}
\newcommand{\eea}{\end{eqnarray}}
\newcommand{\ba}{\begin{array}}
\newcommand{\ea}{\end{array}}

\def\a{_{\alpha}}

\def\1{_{1}}

\begin{document}
\draft
\title{Electronic states in a quantum lens}
\author{Arezky H. Rodr\'{\i}guez$^{(a)}$, C. Trallero-Giner$^{(a)}$,
S.E. Ulloa$^{(b)} $ and J. Mar\'{\i}n-Antu\~{n}a$^{(a)}$}
\address{$^{(a)}$ Department of Theoretical Physics, University of Havana, 10400,\\
Havana, Cuba \\
$^{(b)}$ Department of Physics and Astronomy and CMSS Program, Ohio\\
University, Athens, Ohio 45701-2979 }
\date{18 July 2000}
\maketitle

\begin{abstract}
We present a model to find analytically the electronic states in
self-assembled quantum dots with a truncated spherical cap
(`lens') geometry. A conformal analytical image is designed to
map the quantum dot boundary into a dot with semi-spherical
shape. The Hamiltonian for a carrier confined in the quantum lens
is correspondingly mapped into an equivalent operator and its
eigenvalues and eigenfunctions for the corresponding Dirichlet
problem are analyzed. A modified Rayleigh-Schr\"{o}dinger
perturbation theory is presented to obtain analytical expressions
for the energy levels and wavefunctions as a function of the
spherical cap height $b$ and radius $a$ of the circular cross
section. Calculations for a hard wall confinement potential are
presented, and the effect of decreasing symmetry on the energy
values and eigenfunctions of the lens-shape quantum dot is
studied. As the degeneracies of a semi-circular geometry are
broken for $b\neq a$, our perturbation approach allows tracking
of the split states. Energy states and electronic wavefunctions
with $m=0$ present the most pronounced influence on the reduction
of the lens height. The analytical expressions presented here can
be used to better parameterize the states in realistic
self-assembled quantum dots.
\end{abstract}


%\begin{multicols}{2}

\section{Introduction}

Quantum dots obtained by interrupted growth in strained
semiconductor interfaces are currently under intense study by
many experimental and theoretical groups. \cite{Petroff-gen}
These `self-assembled' quantum dots are mostly dislocation free,
coherent islands of deposited material on the surface of a
different semiconductor. The lattice mismatch from one
semiconductor to the other forces the segregation of material
whenever the epitaxial growth exceeds a critical layer thickness,
resulting in the growth of these so-called Stranski-Krastanov
islands. \cite{zangwill-book} Deposition of material past a
critical thickness, which depends on the two materials used,
results in large arrays of small islands with a rather narrow
size distribution (with size variations well under 10\%), and
typically arranged randomly on the plane (although avoiding
overlapping islands, for the most part). \cite{Petroff-gen} More
recently, some groups are working at producing in-plane ordering
of islands following different approaches, including `nucleation
site engineering' to favor certain locations for self-assembled
dot growth. \cite{Petroff-ordering} These self-assembled quantum
dots have, for the most part, a large area-to-height aspect
ratio. In the case of InAs islands grown on a GaAs surface, the
in-plane diameters are typically less than 30 nm, while their
heights are below 10 nm. \cite{leonard-apl93} Although not
without some controversy, it is generally believed that the InAs
quantum dots on GaAs are shaped like a `lens', characterized by a
spherical cap shape with circular cross section. Upon optical or
other mechanisms of carrier injection, both electrons or holes
have bound well-confined states inside these dots.

In fact, optical experiments on self-assembled dots demonstrate
that these structures provide strong carrier confinement, as
decreasing dot size produces strong blue-shift of the
extremely-narrow luminescence peaks in {\em isolated} dots.
\cite{t3,t6,t7,t11,Jackson} Confinement effects have also been
shown to appear in magneto-capacitance and infrared absorption
experiments by several groups.
\cite{drexler-prl,lorke-prb,gio-preprint,rings} Clear evidence of
electronic shell states and their different degeneracies has been
reported recently, \cite{barbara-prep} and novel applications
such as storage of photoluminescence signals was demonstrated
recently in gate-activated `optical memories.' \cite{Petroff-Sci}

Simplifying the symmetry of the lens as a two-dimensional
harmonic oscillator has been suggested to characterize the level
structure for charge carriers in the dots, and this has proven
useful in the interpretation of experiments.
\cite{PaweldotNature} Nevertheless, as we will show here, a lens
geometry has quite a different level structure and wavefunctions
with subtle symmetries which might be seen in experiments. For
example, we find that as the height of the cap or lens decreases,
there is a larger shift of the wavefunction towards the plane of
the lens (in comparison with the situation in a semi-spherical
geometry). This shift becomes stronger for `flatter' lenses, and
may even give rise to deconfinement of the state towards the
substrate, for finite confinement potential, changing
significantly the oscillator strength of electron-hole
transitions, for example. \cite{Zunger} However, even before that
occurs, the smaller height lens geometries exhibit a different
level structure than the harmonic oscillator, as we show below.
This structure may give rise to different Pauli blocking effects
and transition rules, topics which are the subject of interest in
recent experiments. \cite{ringsNature}

In this paper we rigorously show that a complete set of
wavefunctions can be generated with the correct lens symmetry to
describe the physical properties given by the differential
equation of interest. For the sake of simplicity, we have applied
our calculation here to the single-particle Schr\"{o}dinger
equation.  The incorporation of non-parabolic band dispersion or
many-particle interactions is straightforward (if only a bit
cumbersome). Notice that while our calculation provides an
analytical approach to the understanding of electronic states in
these structures, it also provides an interesting example of a
generalization of perturbation theory for the conformally-mapped
differential operator arising from the Schr\"{o}dinger equation
in the material. Since the conformal transformation used to map
the lens into the semi-spherical geometry is non-trivial, the
resulting differential equation in the transformed space reflects
that complexity. Fortunately, our perturbation approach is robust
for the identified small parameter of the problem, as we will
show.

The remainder of the paper is structured as follows: in section
II we introduce the problem of the spherical cap geometry, its
semi-spherical limit, and the conformal map which connects the two
shapes. In that section, we also describe the perturbation
approach needed to carry out the solution of the appropriately
mapped Schr\"odinger operator, and explore the orthonormalization
conditions of the basis used in the description of the general
problem. In section III we present some examples of the
eigenstates for varying lens size, and analyze their angular and
radial distribution functions. Section IV presents some
discussions and conclusions, while the appendix contains details
of the expansion described in section II.

\section{Dirichlet problem for a quantum lens: Conformal mapping}

The shape of the quantum lens in real $\mbox{\boldmath
$r$}$-space is shown in Fig.\ 1(a). $R_{3}(a,b)$ denotes the
domain of the lens with boundary in $\mbox{\boldmath $r$}$-space
given by a spherical cap of height $b$ and circular cross section
with radius $a$. Assuming a carrier Hamiltonian model with
isotropic band and effective mass $m^{\ast }$, the eigenvalue
problem for a particle confined in the lens is described by the
operator
\begin{equation}
\hat{H}=-\frac{\hbar ^{2}}{2m^{\ast }}\nabla _{\mbox{\boldmath
$r$}}^{2}\hspace*{1cm};\hspace*{1cm}\mbox{\boldmath $r$}\in R_{3}(a,b)\,,
\label{operator}
\end{equation}
obeying the Dirichlet boundary condition $\psi =0$ for
$\mbox{\boldmath $r$} \in L_{3}(a,b)$. $L_{3}(a,b)$ is the
boundary of the $R_{3}(a,b)$ domain. The operator
(\ref{operator}) presents axial symmetry and all functions
defined on $R_{3}(a,b)$ have the property $\phi (\varphi )=\phi
(\varphi +2\pi )$, where $\varphi $ is the axial angle. Hence,
the solution of (\ref {operator}) can be written as
\begin{equation}
\psi (\mbox{\boldmath $r$})=f(\mbox{\boldmath
$\rho$})\frac{\mathop{\rm \mbox{{\large
e}}}\nolimits^{i\;m\;\varphi }}{\sqrt{2\pi
}}\hspace*{1cm};\hspace* {1cm}m=0,\pm 1,\pm 2,...,  \label{2}
\end{equation}
where $\mbox{\boldmath $\rho$}$ is a 2D vector. Correspondingly,
the operator (\ref{operator}) is transformed into an eigenvalue
problem for $f( \mbox{\boldmath $\rho$})$,
\begin{equation}
\left[ \nabla _{\mbox{\boldmath $\rho$}}^{2}+\left(
k_{o}^{2}-\frac{m^{2}-1/4 }{\rho ^{2}\sin ^{2}\theta }\right)
\right] f(\mbox{\boldmath $\rho$})=0
\hspace*{1cm};\hspace*{1cm}\mbox{\boldmath $\rho$}\,\in
\,R_{2}(a,b)\,, \label{operator2}
\end{equation}
where $k_{o}^{2}=2m^{\ast }E/\hbar ^{2}$, $E$ is the energy of the
eigenfunction $f(\mbox{\boldmath $\rho$})$, $\theta $ is the
polar angle, and $R_{2}(a,b)$ is the 2D domain with boundary
$L_{2}(a,b)$ in the $\mbox{\boldmath $\rho$}$-space shown in
Fig.\ 1(b). The Hilbert space where the operator
(\ref{operator2}) is defined corresponds to the set of functions
$f(\mbox{\boldmath $\rho$})\in R_{2}(a,b)$ with boundary condition
$f(\mbox{\boldmath $\rho$})=0$ on $L_{2}(a,b)$.

\subsection{Semi-spherical quantum dot}

For the case of $b=a$, the lens shape of Fig.\ 1(a) has semi-spherical
symmetry. Hence, the Dirichlet problem in (\ref{operator2}) reduces to the
conditions $f(a,\theta )=0$, and $f(\rho ,\pi /2)=0$. The set $\{f_{i}\}$ of
eigenfunctions of (\ref{operator2}) on the $R_{2}$ domain forms an
orthonormal basis with functions which in polar coordinates are given by
products of the associate Legendre polynomials and Bessel functions,
\begin{equation}
f_{n,l}^{(0)}(\rho ,\theta )=\sqrt{\sin \theta} \; \;
\frac{P_{l}^{|m|}(\cos \theta)} {N_{l,m}} \; \;
\frac{J_{l+\frac{1}{2}}(\frac{\textstyle{\mu _{n}^{(l)}}}{
\textstyle{a}}\rho )}{N_{B}} \, ,  \label{3}
\end{equation}
with $l=0,1,2,...$, and the condition $-\;l\leq m\leq l$. The normalization
constants $N_{l,m}$ and $N_{B}$ are as usual given by
\begin{equation}
N_{l,m}=\sqrt{\frac{1}{2l+1}\;\frac{(l+|m|)!}{(l-|m|)!}}\hspace*{5mm};
\hspace*{5mm}N_{B}=\frac{a}{\sqrt{2}}\;J_{l+\frac{1}{2}}^{\prime
}(\mu _{n}^{(l)}) \, ,  \label{4}
\end{equation}
where $J_{q}^{\prime }$ is the derivative of the Bessel function
$J_{q}$, and $\mu _{n}^{(p)}$ is the n-th zero,
$J_{p+\frac{1}{2}}(\mu_n^{(p)})=0$. The eigenvalues are given by
$E_{n,l}=\hbar ^2 \left( \mu _{n}^{(l)}\right) ^{2}/(2m^{\ast
}a^{2})$, and the boundary condition $f^{(0)}(\rho ,\theta =\pi
/2)=0$ restricts the values of the quantum numbers $l$ and $m$ to
fulfill the condition $|l-m|=odd$. According to this condition,
the degeneracy of states $f_{n,l}^{(0)}$ for a given energy
$E_{n,l}$ is equal to $l$ and the ground state corresponds to
$l=1$, $m=0,$ and $n=1$.

\subsection{Quantum lens}

The quantum dot with lens shape corresponds to the more general
case when $b<a$. Here, we need to fulfill Eq.\ (\ref{operator2})
with the Dirichlet condition over the boundary $L_{2}(a,b)$. The
wavefunctions $f_{n,l}^{(0)}$ given by (\ref{3}) are not solution
for the general case, because the problem has no longer the
semi-circular symmetry in $\theta $. The energy number $n$ and
angular momentum $l$ are clearly no longer good quantum numbers
when $b\neq a$ and the m-degeneracy is broken. To obtain an
analytical solution of the problem (\ref{operator2}) it is
convenient to make a conformal mapping to the circular cap with
domain $R_{2}(a,a)$ and boundary $L_{2}(a,a)$. The mapping enables
us to solve the Dirichlet problem for the operator given by
(\ref{operator2}) in a Hilbert space where an orthonormal basis
$\{f_{i}\}$ is known. We transform the quantum lens domain and
boundary into a dot with semi-spherical shape, so that the
circular cap defined by the domain ${\mbox{$\cal Z$}}=x-iz\in
R_{2}(a,b)$, transforms into the semi-circular domain ${\cal
W}=u-iv\in R_{2}(a,a)$. This is accomplished by the
transformation (see Fig.\ 1(b))
\begin{equation}
\mbox{$\cal W$}(\mbox{$\cal Z$})=\frac{2a}{1+\left(
\frac{\textstyle{a-\mbox{$\cal Z$}}}{\textstyle{a+\mbox{$\cal
Z$}}}\right) ^{\textstyle{\alpha
}}}-a\hspace*{5mm};\hspace*{5mm}\alpha = \frac{\pi /4}{\arctan
(b/a)}\,.  \label{transformacion}
\end{equation}
In the complex plane $\mbox{$\cal W$}$, we have the parameter
equations: $u=\rho \sin \theta ,$ $v=\rho \cos \theta ,$ $0<\rho
<a,$ $0<\theta <\pi /2.$ The eigenvalue problem (\ref{operator2})
is thus transformed by this conformal mapping into the problem
\begin{equation}
\nabla _{(u,v)}^{2}F(u,v)+\mbox{${\cal J}\a$}(u,v)\left(
k^{2}-\frac{m^{2}-1/4}{\mbox{${\cal X}\a$}^{2}(u,v)}\right)
F(u,v)=0\;;\;(u,v)\in R_{2}(a,a),  \label{pp}
\end{equation}
with the boundary condition,
\begin{equation}
\left. F(u,v)\right. |_{\textstyle{(u,v)\in L_{2}(a,a)}}=0\,.  \label{ppp}
\end{equation}
The functions $\mbox{${\cal J}\a$}(u,v)=\left|
\frac{\textstyle{\mbox{d}\mbox{$\cal
Z$}}}{\textstyle{\mbox{d}\mbox{$\cal W$}}}\right| ^{2}$ (the
Jacobian) and $\mbox{${\cal J}\a$}/\mbox{${\cal X}\a$}^{2}$ are
given in the Appendix, and are mathematical objects which contain
the information of the lens geometry, where the subscript $\alpha
$ is given in (\ref {transformacion}). It should be noted that
$\alpha \geq 1$, since $b\leq a$, and for $\alpha =1$ the
Jacobian $\mbox{${\cal J}\a$}$ reduces to $1$, while ${\cal
J}_{\alpha }/{\cal X}_{\alpha }^{2}$ reduces to $1/u^{2}=1/\left(
\rho \sin \theta \right) ^{2}$.

The Hilbert space on which the operator (\ref{pp}) is defined
must fulfill the Dirichlet boundary conditions indicated in
(\ref{ppp}). A set of functions that fulfill these boundary
conditions are the functions $f_{n,l}^{(0)}$ given in (\ref{3}),
and represent a complete set of orthonormal eigenfunctions for
the operator in (\ref{pp}). The solution $F(u,v)$ for a given $m$
can be expanded in term of the set $\{f_{n,l}^{(0)}\} $ such that
\begin{equation}
F=\sum_{n,l}C_{n,l}f_{n,l}^{(0)}(\mbox{\boldmath $\rho$})\,,  \label{5}
\end{equation}
where \mbox{\boldmath $\rho$}$=(\rho ,\theta )$ is here the parameterization
of $(u,v)$, and the functions $f_{n,l}^{(0)}$ are restricted to the
condition $|l-m|=odd$. The coefficients $C_{n,l}$ have to be determined to
satisfy the full operator; a perturbation procedure to accomplish this is
described in section \ref{coeffs} below.

\subsection{Orthogonality and completeness}

Equation (\ref{pp}) can be cast in operator form as
\begin{equation}
\mbox{\boldmath $\hat K$}F=\mbox{${\cal J}\a$}k^{2}F,  \label{6}
\end{equation}
where $\mbox{\boldmath $\hat K$}$ involves the Laplace operator, $-\nabla
_{(u,v)}^{2}$, and the term $(m^{2}-1/4)\mbox{${\cal
J}\a$}/\mbox{${\cal X}\a$}^{2}(u,v)$. Equation (\ref{6}) is an eigenvalue
problem for the dimensionless energies $k_{N}^{2}(m)$, where $N$ is a
generic label for the different eigenstates $F_{N}$ of (\ref{6}). Let us now
suppose that $N$ and $M$ correspond to different eigenvalues of (\ref{6}).
From the above equation it follows that
\begin{equation}
F_{M}^{\ast }\mbox{\boldmath $\hat K$}F_{N}-F_{N}\mbox{\boldmath
$\hat K$}^{\ast }F_{M}^{\ast }=\mbox{${\cal
J}\a$}(k_{N}^{2}-k_{M}^{2})F_{M}^{\ast }F_{N}\,.  \label{7}
\end{equation}
Making the integration over $\mbox{\boldmath $\rho$}(u,v)$ in the $R_{2}(a,a)
$ domain, we obtain that
\begin{equation}
\int_{R_{2}(a,a)}\left[ -F_{M}^{\ast }\nabla
_{(u,v)}^{2}F_{N}+F_{N}\nabla _{(u,v)}^{2}F_{M}^{\ast }\right]
\mbox{d}^{2}\rho
=(k_{N}^{2}-k_{M}^{2})\int_{R_{2}(a,a)}F_{M}^{\ast
}F_{N}\mbox{${\cal J}\a$} \mbox{d}^{2}\rho \,.  \label{8}
\end{equation}
Integration by parts gives us
\begin{equation}
(k_{N}^{2}-k_{M}^{2})\int_{R_{2}(a,a)}\mbox{${\cal
J}\a$}F_{M}^{\ast }F_{N}\;\mbox{d}^{2}\rho =\left( F_{N}\nabla F_{M}^{\ast
}-F_{M}^{\ast }\nabla F_{N}\right) |_{(u,v)\in L_{2}(a,a)}\,,  \label{9}
\end{equation}
which due to the boundary condition (\ref{ppp}), it is reduced to
\begin{equation}
(k_{N}^{2}-k_{M}^{2})\int_{R_{2}(a,a)}\mbox{${\cal
J}\a$}F_{M}^{\ast }F_{N}\;\mbox{d}^{2}\rho =0\,.  \label{10}
\end{equation}
For $N\neq M$, condition (\ref{10}) represents the orthogonality
property of the eigenfunction set $\{F_{N}\}$, where
$\mbox{${\cal J}\a$}$ is clearly the {\it weighting factor} of
the eigenproblem (\ref{pp}). Moreover,  the operator
$\mbox{\boldmath $\hat K$}$ is Hermitian, ensuring that  the
solution of the present problem is described by means of a
complete orthonormal basis of eigenfunctions $\{F_{N}\}$ obeying
the condition
\begin{equation}
\int_{R_{2}(a,a)}\mbox{${\cal J}\a$}F_{M}^{\ast }F_{N}\;\mbox{d}^{2}\rho
=\delta _{N,M}\,.  \label{11}
\end{equation}

\subsection{Perturbation theory}

\label{coeffs}

The coefficients $C_{n,l}$ in (\ref{5}) and the eigenvalues $k^{2}$ can be
obtained by perturbation theory if $b\approx a\;(\alpha \rightarrow 1)$. In
this case, the lens cap represents a perturbation from the semi-spherical
geometry. In other words, the operator (\ref{pp}) can be rewritten in the
form
\begin{equation}
\left( H_{o}+H_{p}\right) F(u,v)=0 \, ,  \label{12}
\end{equation}
with
\begin{eqnarray}
H_{o}(u,v) &=&\nabla _{(u,v)}^{2}+\left( k^{2}-\frac{m^{2}-1/4}{u^{2}}
\right) ,  \label{13} \\
H_{p}(u,v) &=& k^{2} \Bigl( \mbox{${\cal J}\a$}(u,v)-1 \Bigl)
-(m^{2}-1/4)\left( \frac{\mbox{${\cal J}\a$}(u,v)}{\mbox{${\cal
X}\a$} ^{2}(u,v)}-\frac{1}{u^{2}}\right) .  \label{14}
\end{eqnarray}
The operator $H_{p}$ vanishes when $\alpha \rightarrow
1\;(b\rightarrow a)$ and it can be considered as a small
perturbation operator, when the height $b $ is close to the
radius $a$. The set $\{f_{n,l}^{(0)}\}$ given by Eq.\ (\ref {3})
are the eigenfunctions of the Hamiltonian $H_{o}$ in the $
\mbox{$\cal W$}$-space and form an orthonormal basis on the
$R_{2}(a,a)$ domain. In order to find the solution of (\ref{12})
as a function of the ratio $b/a$, we will develop a modified
Rayleigh-Schr\"{o}dinger perturbation theory. We note that the
perturbation Hamiltonian $H_{p}$ depends on the eigenvalue
$k^{2}$, and as such requires a somewhat different approach.
Substituting Eq.\ (\ref{5}) in (\ref{12}) we obtain
\begin{eqnarray}
\left[ \left( k^{2}-k_{o}^{2}\right) \right. &+&\left. \langle \;
n,l\;|\;H_{p}(k^{2})\;|\;n,l \; \rangle \right] C_{n,l} \; +  \nonumber \\
&+&\;\sum_{n^{\prime },l^{\prime }\neq n,l} \langle
\;n,l\;|\;H_{p}(k^{2})\;|\;n^{\prime },l^{\prime }\; \rangle \; C_{n^{\prime
},l^{\prime }}=0.  \label{15}
\end{eqnarray}
The states of $H_{o}$ are degenerate on the quantum number $m$.
Nevertheless, according to (\ref{2}),
\begin{equation}
\langle n,l,m\;|\;H_{p}\;|\;n^{\prime },l^{\prime },m^{\prime } \rangle
\;=\; \langle n,l\;|\;H_{p}\;|\;n^{\prime },l^{\prime } \rangle \;\delta
_{m,\;m^{\prime }},  \label{16}
\end{equation}
so that the matrix elements are diagonal on $m$ and we can
develop a perturbation theory in the absence of degeneracy. We
represent the coefficients $C_{n,l}$ and its eigenvalues $k^{2}$
in a power series of the small parameter $\lambda =\alpha
^{-1}-1$ (which arises naturally from the expressions in the
Appendix). We obtain up to first order in $\lambda$ an expression
for the wavefunctions given by
\begin{eqnarray}
F_{N,m}(\rho ,\theta ) &=&f_{n,l}^{(0)}(\rho ,\theta )+\lambda \left[ -\frac{
1}{2}\left\langle n,l\;\left| \left( \frac{\partial \mbox{${\cal J}\a$}}{
\partial \lambda }\right) _{\lambda =0}\right| \;n,l\right\rangle \right.
f_{n,l}^{(0)}(\rho ,\theta )  \nonumber \\
&&+\sum_{
\begin{tabular}{ccc}
$n^{\prime },l^{\prime }$ & $\neq $ & $n,l$
\end{tabular}
}f_{n^{\prime },l^{\prime }}^{(0)}(\rho ,\theta )\times \frac{1}{
k_{o}^{2}(n^{\prime },l^{\prime })-k_{o}^{2}(n,l)}\;\times  \nonumber \\
&&\left. \left\langle n^{\prime },l^{\prime }\left|
k_{o}^{2}(n,l)\left( \frac{\partial \mbox{${\cal J}\a$}}{\partial
\lambda }\right) _{\lambda =0}-(m^{2}-1/4)\frac{\partial
}{\partial \lambda }\left( \frac{\mbox{${\cal
J}\a$}}{\mbox{${\cal X}\a$}^{2}}\right) _{\lambda =0}\right|
n,l\right\rangle \right] \, .  \label{17}
\end{eqnarray}
We find for the eigenvalues up to second order, that
\begin{equation}
k^{2}(N,m)=k_{o}^{2}(n,l)+\lambda \;k_{1}^{2}(N,m)+\lambda
^{2}\;k_{2}^{2}(N,m),  \label{18}
\end{equation}
where
\begin{eqnarray}
k_{1}^{2}(N,m) &=&-\;\left\langle n,l\left| \;k_{o}^{2}(n,l)\left( \frac{
\partial \mbox{${\cal J}\a$}}{\partial \lambda }\right) _{\lambda
=0}-(m^{2}-1/4)\frac{\partial }{\partial \lambda }\left( \frac{\mbox{${\cal
J}\a$}}{\mbox{${\cal X}\a$}^{2}}\right) _{\lambda =0}\right|
n,l\right\rangle ,  \label{19} \\
&&\   \nonumber \\
k_{2}^{2}(N,m) &=&-\; \mathrel{\mathop{\sum }\limits_{
\begin{tabular}{ccc} $n^{\prime },l^{\prime }$ & $\neq $ & $n,l$
\end{tabular} }} \frac{
\left\langle n^{\prime},l^{\prime}\left| k_{o}^{2}(n,l)\left(
\frac{ \textstyle{\partial} \mbox{${\cal
J}\a$}}{\textstyle{\partial \lambda}}\right) _{\lambda =0}-(m^{2}-1/4)\frac{%
\textstyle{\partial}}{\textstyle{\partial \lambda}}\left(
\frac{\mbox{${\cal J}\a$}}{\mbox{${\cal X}\a$}^{2}}\right)
_{\lambda =0}\right| n,l\right\rangle^{2} }{
k_{o}^{2}(n^{\prime },l^{\prime }) - k_{o}^{2}(n,l)}  \nonumber \\
&&\ -\frac{1}{2}\left\langle n,l\left| k_{1}^{2}(N,m)\left(
\frac{\partial \mbox{${\cal J}\a$}}{\partial \lambda }\right)
_{\lambda =0}\;+\right. \right. k_{o}^{2}(n,l)\left(
\frac{\partial ^{2}\mbox{${\cal J}\a$}}{
\partial \lambda ^{2}}\right) _{\lambda =0}  \nonumber \\
&&\;\ \ \ \ \ \ \ \ \ \ \ \ \ \ \ \ \ \ \ \left. \left.
-(m^{2}-1/4)\frac{
\partial ^{2}}{\partial \lambda ^{2}}\left( \frac{\mbox{${\cal J}\a$}}{
\mbox{${\cal X}\a$}^{2}}\right) _{\lambda =0}\right| n,l\right\rangle \, .
\label{20}
\end{eqnarray}
The different factors included in the geometric perturbation,
\begin{equation}
\left( \frac{\partial \mbox{${\cal J}\a$}}{\partial \lambda
}\right) _{\lambda =0},\;\left( \frac{\partial ^{2}\mbox{${\cal
J}\a$}}{\partial \lambda ^{2}}\right) _{\lambda
=0},\frac{\partial }{
\partial \lambda }\left( \frac{\mbox{${\cal
J}\a$}}{\mbox{${\cal X}\a$}^{2}}\right) _{\lambda =0},\;
\frac{\partial ^{2} }{\partial \lambda ^{2}}\left(
\frac{\mbox{${\cal J}\a$}}{\mbox{${\cal X}\a$}^{2}}\right)
_{\lambda =0}  \label{chorizo}
\end{equation}
are also given in the Appendix. In the framework of the infinite
confinement model, the parameter dependence of Eqs.\
(\ref{17})-(\ref{20}) is known, since the matrix elements play
the role of constants and need to be evaluated only once. Note
also that the above expressions are not the same as those found
in a typical perturbation theory, as the difference arising in
the $k^{2}$ term depends on the perturbation Hamiltonian $H_{p}$,
which itself depends on the parameter $\lambda$.

\section{The eigenvalues and wavefunctions}

Figure 2(a) shows the first 13 energy levels in units of $E_{o}=\hbar
^{2}/(2m^{\ast }a^{2})$ for a quantum lens calculated by perturbation theory
up to the second order in $\lambda = \alpha^{-1} -1$, as a function of the
ratio $b/a$. The different eigenvalue curves are labeled by the quantum
numbers $(N,m)$. The semi-sphere case $(b/a=1)$ is indicated by the limiting
value on the right vertical axis in each panel. One can see the breaking of
degeneracy in the quantum number $m$, and the strong deviation from the
semi-spherical case, as the ratio $b/a$ decreases. The lower levels exhibit
a weaker dependence on the decreasing $b/a$ ratio, while the upper levels
are strongly deviated from the semi-spherical case. In Fig.\ 2(b) the first
5 energy levels calculated up to first (dotted lines) and second order
(solid lines) perturbation on the parameter $\lambda $ are compared in the
range $0.4\leq b/a\leq 1$. It can be seen that a strong deviation is present
for the higher excited levels ($N=3$ and 4), while for $N=1$ and 2 the
obtained results using Eq.\ (\ref{19}) (first order perturbation theory)
give a deviation smaller than $1\%$ in comparison with those using Eq.\ (\ref
{20}) (second order perturbation results).

The radial and angular probability density function in a given state $N,m$
are defined as:
\begin{equation}
\ P_{N,m}(\rho )=\int\limits_{0}^{\pi /2}\left| F_{N,m}(\rho
,\theta )\right| ^{2}\rho \;{\rm d}\theta ,\ \text{\ \ \ \
}P_{N,m}(\theta )=\int\limits_{0}^{a}\left| F_{N,m}(\rho ,\theta
)\right| ^{2}\rho \;{\rm d} \rho \,.  \label{21}
\end{equation}
The functions $P(\rho )$ and $P(\theta )$ are obtained up to first order
perturbation theory on $\lambda $ according to the equations
\begin{eqnarray}
P_{N,m}(\rho ) &=&\rho \left( \frac{J_{l+1/2}(\mu _{n}^{(l)}\rho
/a)}{N_{B}} \right) ^{2}\left[ 1+\lambda \left( -\left\langle
n,l\left| \left( \frac{
\partial \mbox{${\cal J}\a$}}{\partial \lambda }\right) _{\lambda =0}\right|
n,l\right\rangle \right. \right. \;+  \nonumber \\
&+&\;\left. \left. \int_{0}^{\pi /2}\mbox{d}\theta \sin \theta
\left( \frac{
\partial \mbox{${\cal J}\a$}}{\partial \lambda }\right) _{\lambda =0}\left(
\frac{P_{l}^{|m|}(\cos \theta )}{N_{l,m}}\right) ^{2}\right) \right]
\;;\;\rho \in (0,a)  \label{22}
\end{eqnarray}
and
\begin{eqnarray}
P_{N,m}(\theta ) &=&\sin \theta \left( \frac{P_{l}^{|m|}(\cos
\theta )}{ N_{l,m}}\right) ^{2}\left[ 1+\lambda \left(
-\left\langle n,l\left| \left( \frac{\partial \mbox{${\cal
J}\a$}}{\partial \lambda }\right) _{\lambda =0}\right|
n,l\right\rangle
\right. \right. \;+  \nonumber \\
&+&\;\left. \left. \int_{0}^{a}\mbox{d}\rho \;\rho \left(
\frac{\partial \mbox{${\cal J}\a$}}{\partial \lambda }\right)
_{\lambda =0}\left( \frac{ J_{l+1/2}(\mu _{n}^{(l)}\rho
/a)}{N_{B}}\right) ^{2}\right) \right] \;;\;\theta \in (0,\pi
/2).  \label{23}
\end{eqnarray}
Figure 3(a) and (b) show the radial and angular probabilities
given by Eqs.\ (\ref{22}) and (\ref{23}), respectively. In both
cases the levels considered are ($N,m)=$ $(1,0)$, $(4,0)$,
$(8,0),$ and $(8,2)$. Solid lines represent the semi-spherical
case ($b/a=1$), while the quantum cap with $b/a=0.509$ is shown
by dotted lines, and illustrates the departure from the
semi-spherical case for decreasing $b/a$ ratio. In the case of
the radial probability, the deviation observed when the ratio
$b/a$ decreases is relatively small, in comparison with the
semi-spherical cap. In contrast, the angular probability shows a
rather strong deviation as a function of $b/a$, as the maximum
probability is shifted towards $\theta =\pi /2$; that is, the
carrier is located more towards the plane $v=0$, as the cap
height decreases (see Fig.\ 1(b)). For all states, the maximum of
$P(\rho )$ is smoothly shifted towards $\rho =0$, except for
level $(8,2)$. The different behavior observed for the radial and
angular probability densities can be seen as arising from the
fact that the geometry of the quantum dot is essentially
decreasing in radius and not in angle, as $b/a$ decreases. The
radial probability is obtained by integration along the angle in
all directions $(0<\theta <\pi /2) $, taking into account the
angular contribution for a given $\rho $ and certain geometry. On
the other hand, the angular probability is calculated by taking
the integration along the radius where the change of geometry is
more important. Hence, one can say that the angular probability
density $P(\theta )$ contains more information about the changing
cap geometry than the function $P(\rho )$, as function of $b/a$.

Finally, a test for the viability to obtain the wavefunctions and
energies by the perturbation method, is given by the ratio
between the matrix element \ of the perturbation Hamiltonian
$\left\langle n,l|H_{p}(\mbox{${\cal J}\a$},m)|n,l\right\rangle $
with respect to the non-perturbed dimensionless energy
$k_{o}^{2},$
\begin{equation}
\Delta _{N,m}\equiv \frac{\left| \left\langle
n,l|H_{p}(\mbox{${\cal J}\a$} ,m)|n,l\right\rangle \right|
}{k_{o}^{2}(n,l)}\,.  \label{24}
\end{equation}
The parameter $\Delta _{N,m}$ was calculated for the levels of
Fig.\ 2 and is shown in Fig.\ 4 as a function of $b/a$. A
necessary (but not sufficient) condition for the perturbation
theory to be valid is that $\Delta _{N,m}$ must be less than the
unity, and this criterion is fulfilled for the range $0.4<b/a<$
$1$. Another more restrictive condition for the applicability of
this method is that $\left\langle n,l|H_{p}(\mbox{${\cal
J}\a$},m)|n^{\prime },l^{\prime }\right\rangle <\left|
k_{o}^{2}(n,l)-k_{o}^{2}(n^{\prime },l^{\prime })\right| $,
saying that the differences between non-perturbed dimensionless
energy states need to be larger than the matrix elements of the
perturbation. A more complete, non-perturbation, solution of Eq.\
(\ref{pp}) is needed to completely assess the validity of the
perturbation approach. However, one can be confident that for
small $\lambda$ values, the eigenstates and eigenfunctions found
represent an accurate solution to the problem, as the
perturbation and the small parameter is well defined, and the
procedure robust.

\section{Conclusions}

We have presented a formal and systematic conformal analytical
map model to describe quantum dots with lens geometry and
circular cross section arising in the growth of low dimensional
semiconductor systems. The reported transformation can be extended
straightforward to different physical-mathematical models as
electronic states and phonon modes fulfilling a certain
differential equation. The proposed conformal image maps the lens
boundary into a dot with semi-spherical shape, allowing one to
obtain a complete set of orthonormal functions to characterize the
physical problem keeping the full lens symmetry. We have applied
the formalism to the eigenvalue and eigenfunction of the
Schr\"{o}dinger problem in a spherical cap geometry. The conformal
mapping of the equation allows a modified but well defined
Rayleigh-Schr\"{o}dinger perturbation approach, where the cap
height to in-plane radius is used to define the small parameter
of the theory.

We find that the wavefunctions are strongly shifted towards the
flat face, as the height of the lens decreases, while the radial
dependence is not affected as much. This change in the
wavefunctions is interesting on its own, as it reflects the
changes produced by the appropriate operator after the conformal
mapping. Moreover, these changes may have important consequences
for the different electronic and optical properties of
self-assembled quantum dots. We are currently studying those
properties and will report our findings in the future. The
reported energy dependence on cap height to in-plane radius and
semiconductors parameters can be useful to characterize the
geometrical dimensions of these novel semiconductor
nanostructures.

\acknowledgments
SEU acknowledges partial support by the US DOE grant no.\ DE-FG02-91ER45334.

\section*{appendix}

The Jacobian of the transformation $\mbox{$\cal W$}(\mbox{$\cal
Z$})$ is given by

\begin{equation}
\mbox{${\cal J}\a$}(r,\theta )=\frac{16\;(1/\alpha
)^{2}}{R^{1-1/\alpha } \left[ f_{+}^{1/\alpha }+f_{-}^{1/\alpha
}+2R^{1/2\alpha }\cos (\phi /\alpha )\right] ^{2}} \, ,
\eqnum{A1}  \label{A1}
\end{equation}
and the term $\mbox{${\cal J}\a$}/\mbox{${\cal X}\a$}^{2}$ can be cast as
\begin{equation}
\frac{\mbox{${\cal J}\a$}(r,\theta )}{\mbox{${\cal
X}\a$}^{2}(r,\theta )}= \frac{16\;(1/\alpha )^{2}}{R^{1-1/\alpha }\left[
f_{+}^{1/\alpha }-f_{-}^{1/\alpha }\right] ^{2}} \, . \hspace*{5mm}
\eqnum{A2}  \label{A2}
\end{equation}
In the above equations we have defined

\begin{equation}
r=\rho /a\hspace*{5mm};\hspace*{5mm}f_{\pm }=1+r^{2}\pm 2r\sin \theta
\hspace *{5mm};\hspace*{5mm}R=f_{+}\;f_{-},  \eqnum{A3}  \label{A3}
\end{equation}
and
\begin{equation}
\phi =\left\{
\begin{tabular}{cc}
$\mbox{arctg}\left( \frac{\textstyle{2r\cos \theta }}{\textstyle{1-r^{2}}}
\right) $ & $;r<1$ \\
$\pi /2$ & $;r=1$ \, .
\end{tabular}
\right.  \eqnum{A4}  \label{A4}
\end{equation}
From Eqs.\ (\ref{A1}) and (\ref{A2}), it follows that

\begin{equation}
{\cal J}_{\alpha =1}(r,\theta )\equiv 1\text{
};\hspace*{5mm}\left( \frac{ \mbox{${\cal J}\a$}(r,\theta
)}{\mbox{${\cal X}\a$}^{2}(r,\theta )}\right) _{\alpha =1}\equiv
\frac{1}{r^{2}\sin ^{2}\theta } \, ,  \eqnum{A5} \label{A5}
\end{equation}
as one would expect.

Finally, the geometric perturbation factors in Eq.\ (\ref{chorizo}) are
given by
\begin{eqnarray}
\left( \frac{\partial \mbox{${\cal J}\a$}(r,\theta )}{\partial \lambda }
\right) _{\lambda =0} &=&2-r\sin \theta \ln (f_{+}/f_{-})+2\phi r\cos \theta
\, ,  \eqnum{A6}  \label{A6} \\
&&  \nonumber \\
\frac{\partial }{\partial \lambda }\left( \frac{\mbox{${\cal
J}\a$}(r,\theta )}{\mbox{${\cal X}\a$}^{2}(r,\theta )}\right) _{\lambda =0}
&=&\frac{4r\sin \theta -(1+r^{2})\ln (f_{+}/f_{-})}{2r^{3}\sin ^{3}\theta }
\, ,  \eqnum{A7}  \label{A7} \\
&&  \nonumber
\end{eqnarray}

\begin{eqnarray}
\left( \frac{\partial ^{2}\mbox{${\cal J}\a$}(r,\theta
)}{\partial \lambda ^{2}}\right) _{\lambda =0}
&=&\frac{1}{2}\left( \frac{\partial \mbox{${\cal J}\a$}}{\partial
\lambda }\right) _{\lambda =0}\left[ 3\left( \frac{\partial
\mbox{${\cal J}\a$}}{\partial \lambda }\right) _{\lambda
=0}-8\right] -\frac{
(1+r^{2})}{4}\ln ^{2}(f_{+}/f_{-})  \nonumber \\
&&  \nonumber \\
&& + \phi ^{2}(1-r^{2})+2\ln ^{2}R \, ,  \eqnum{A8} \\
&&  \nonumber \\
\frac{\partial ^{2}}{\partial \lambda ^{2}}\left( \frac{\mbox{${\cal J}\a$}
(r,\theta )}{\mbox{${\cal
X}\a$}^{2}(r,\theta )}\right) _{\lambda =0} &=& \frac{(4r\sin \theta
)^{2}-16r\sin \theta (1+r^{2})\ln (f_{+}/f_{-})}{ 8r^{4}\sin ^{4}\theta }
\nonumber \\
&&  \nonumber \\
&&+\frac{(2(1+r^{2})^{2}+R) \ln ^{2}(f_{+}/f_{-})}{8r^{4}\sin ^{4}\theta }
\, .  \eqnum{A9}
\end{eqnarray}

\begin{figure}[tbp]
%\centerline{\epsfxsize=7cm \epsfbox{fig1.eps}}
 \caption{(a) Lens
cap of height $b$ and radius $a$. (b) Conformal mapping from
$\mbox{$\cal Z$}$ to $\mbox{$\cal W$}$.} \label{fig1}
\end{figure}

\begin{figure}[tbp]
%\centerline{\epsfxsize=7cm \epsfbox{fig2a.eps}}
%\centerline{\epsfxsize=7cm \epsfbox{fig2b.eps}}
 \caption{Energy
levels $E_N,_m(b/a)$, labeled by $(N,m)$, for a quantum lens as a
function of the ratio $b/a$. The energies are given in units of
$E_0=2m^*E/\hbar ^{2}$. (a) The first 13 energy levels calculated
up to second order in perturbation theory. (b) Comparison between
results calculated up to first (dotted lines) and second order
(solid) perturbation theory for the first five levels. }
\label{fig2}
\end{figure}

\begin{figure}[tbp]
%\centerline{\epsfxsize=7cm \epsfbox{fig3a.eps}}
%\centerline{\epsfxsize=7cm \epsfbox{fig3b.eps}}
 \caption{(a) The radial probability density in a quantum lens
$P_{N,m}$($\protect\rho$), in units of $a^{-1}$, for different
electronic states $(N,m)$, and as function of the dimensionless
coordinate $r=\protect\rho /a$. (b) Angular probability density
$P_{N,m}$($\protect\theta$) as function of the angle
$\protect\theta$. Two values of the ratio $b/a$ are considered:
$b/a=1$ (solid lines), and $b/a=0.509 $ (dotted lines). The
calculations were based on the perturbation theory described by
Eqs.\ (\ref{22}) and (\ref{23})). Different states $(N,m)=$
(1,0), (4,0), (8,0), and (8,2) are indicated.} \label{fig3}
\end{figure}

\begin{figure}[tbp]
%\centerline{\epsfxsize=7cm \epsfbox{fig4.eps}}
 \caption{Ratio between the matrix element $\Delta$ and the energy
$k_o^{2}(n,l)$ with respect to the ratio $b/a$ of the quantum
lens. For all $b/a$ between $0.4$ and $1$ the parameter $\Delta$
is small.} \label{fig4}
\end{figure}

%\end{multicols}

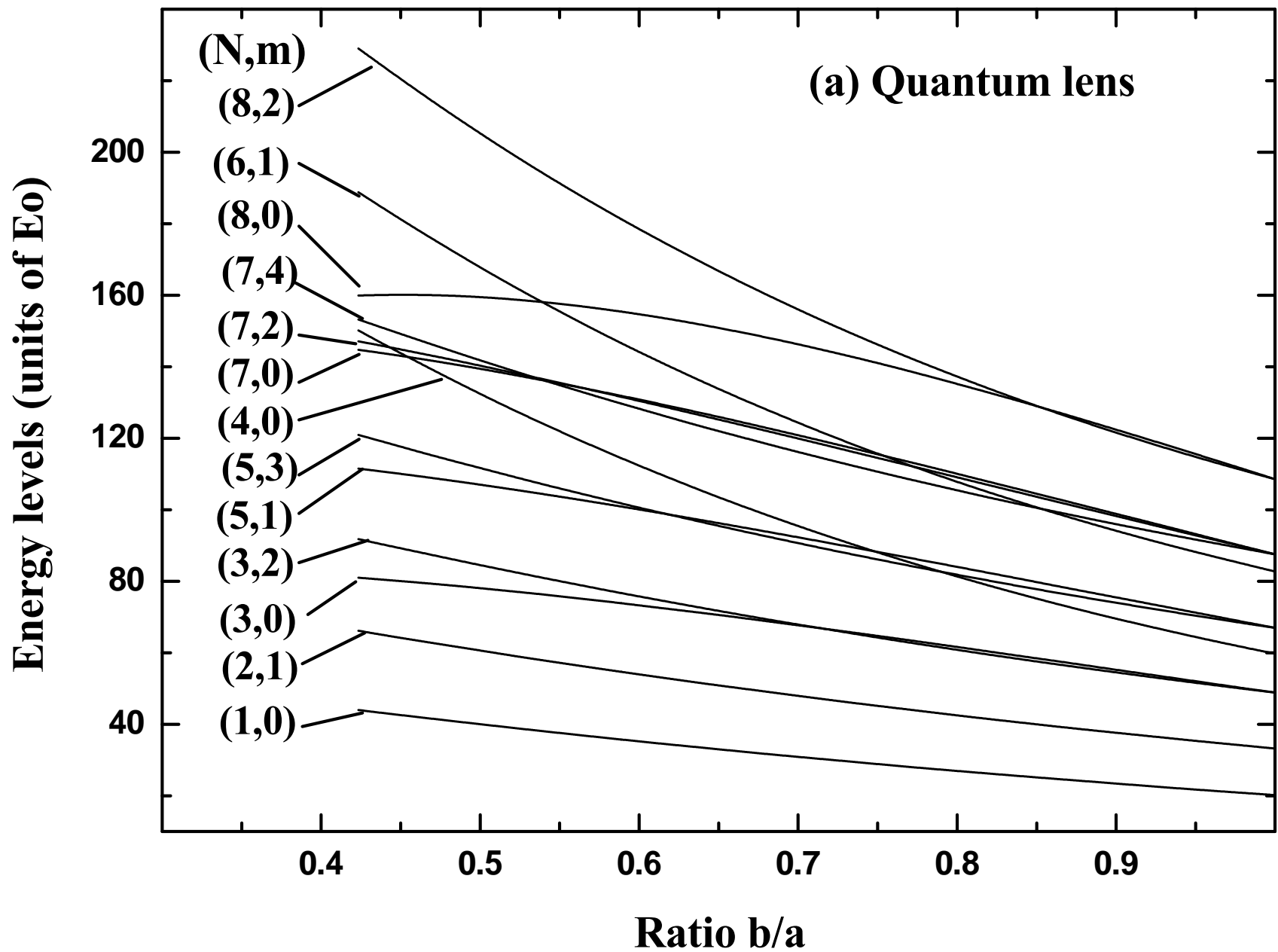
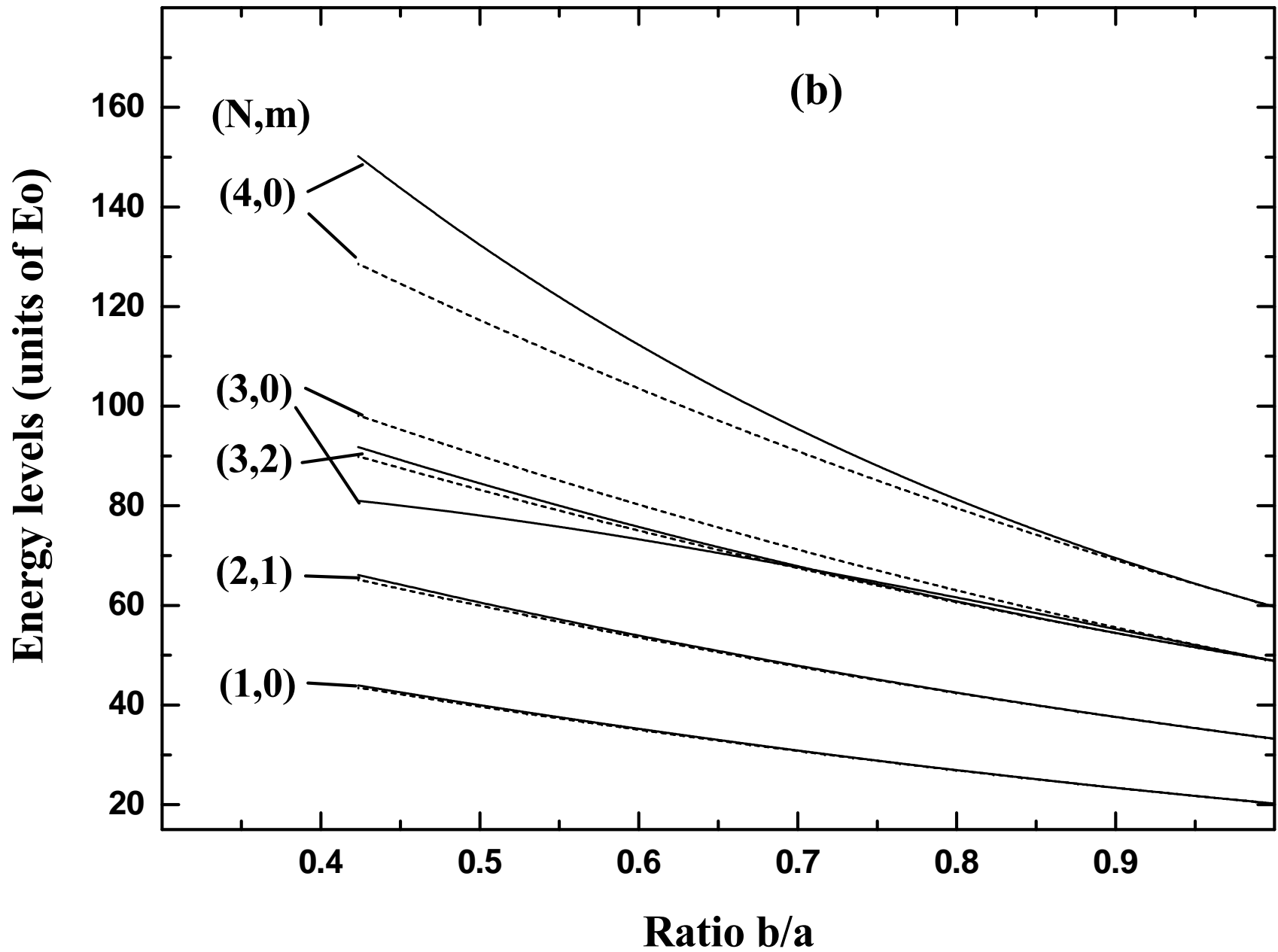
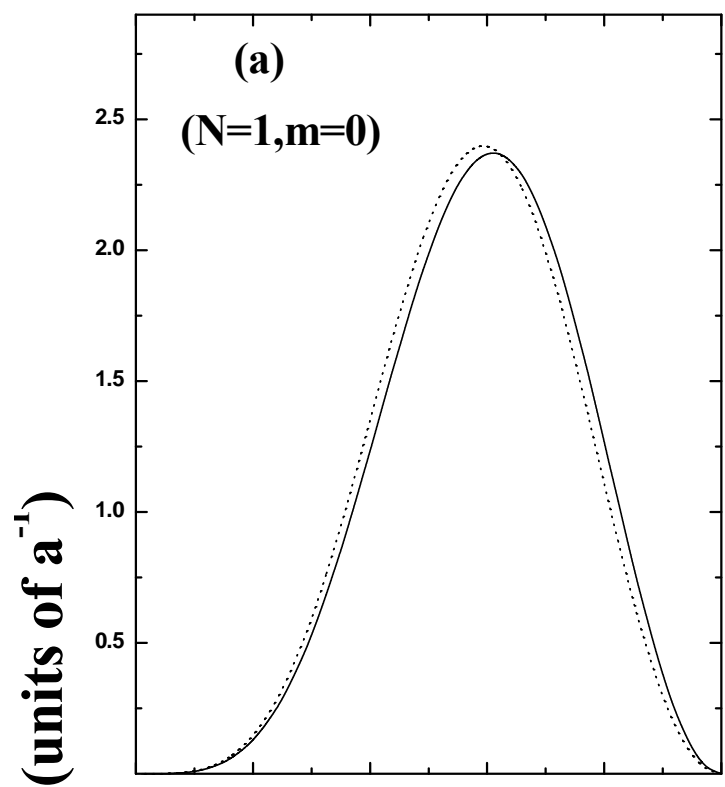
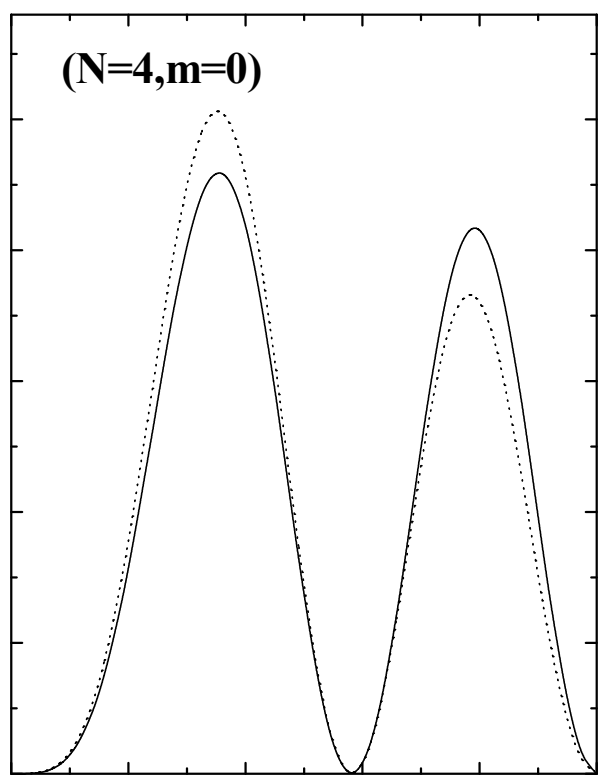
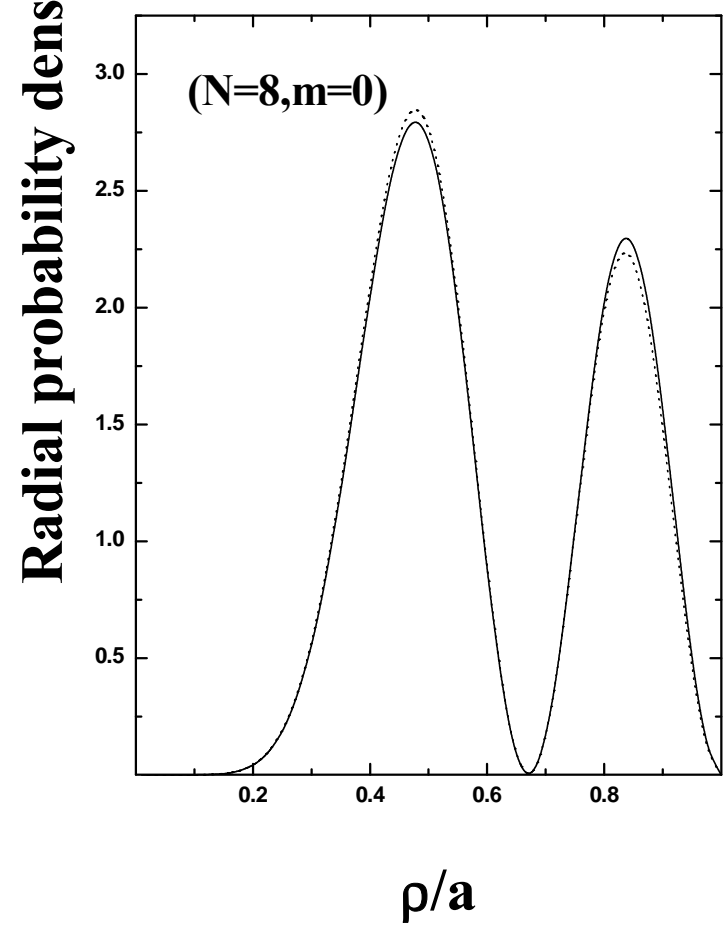
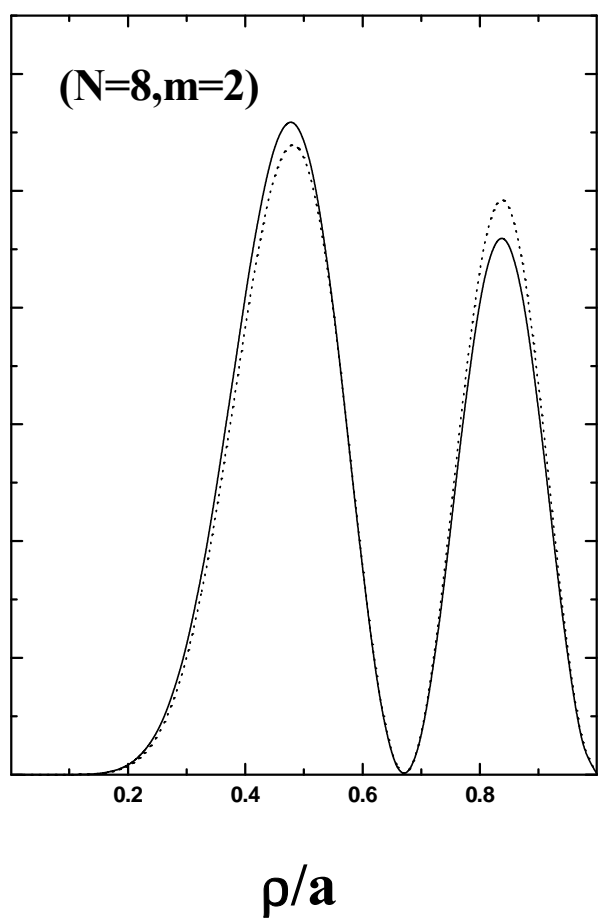
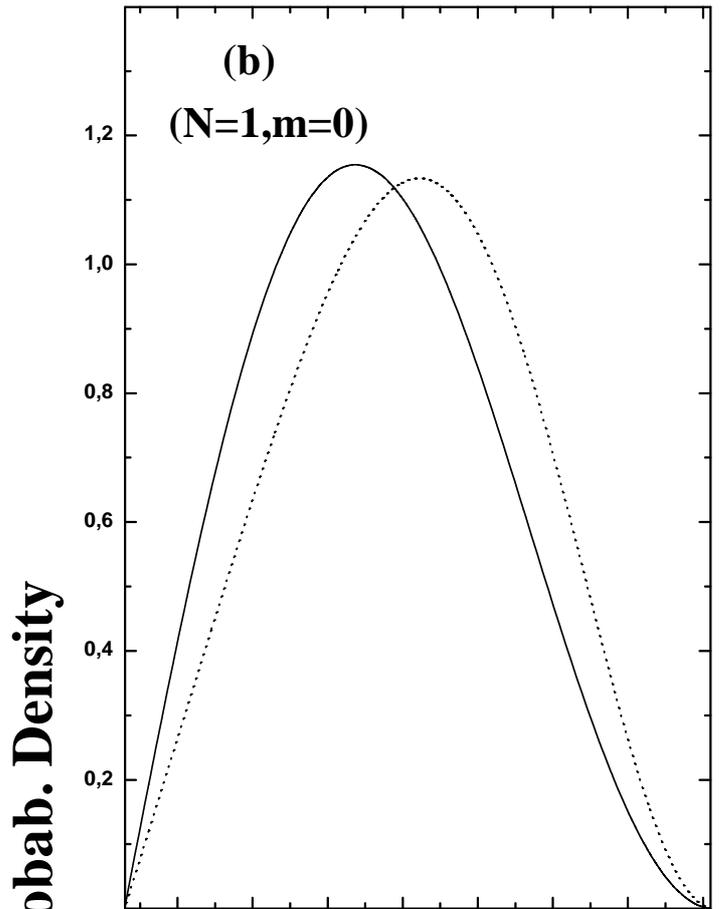
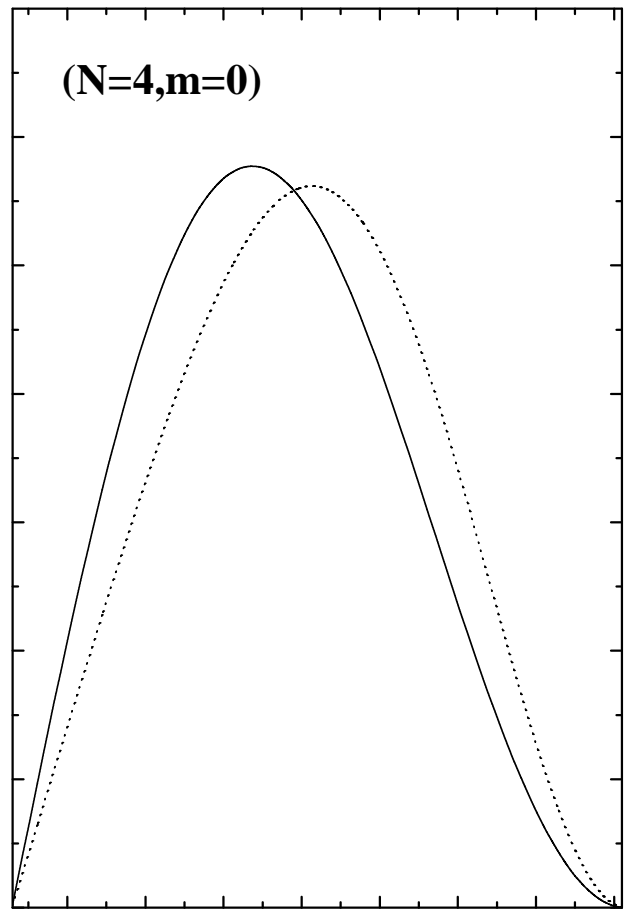
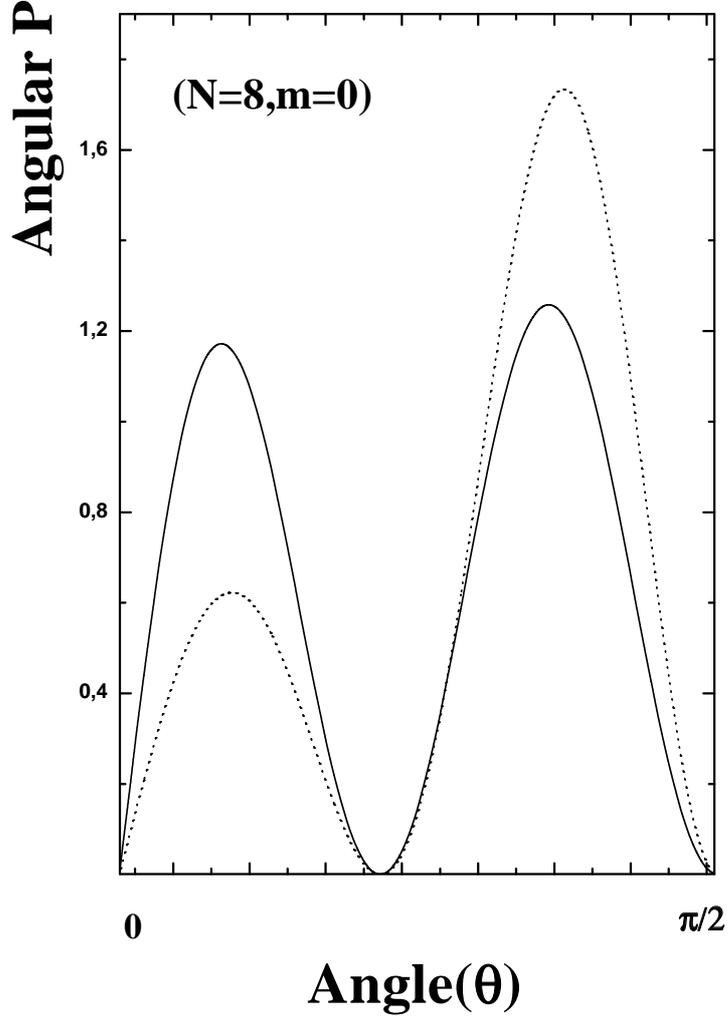
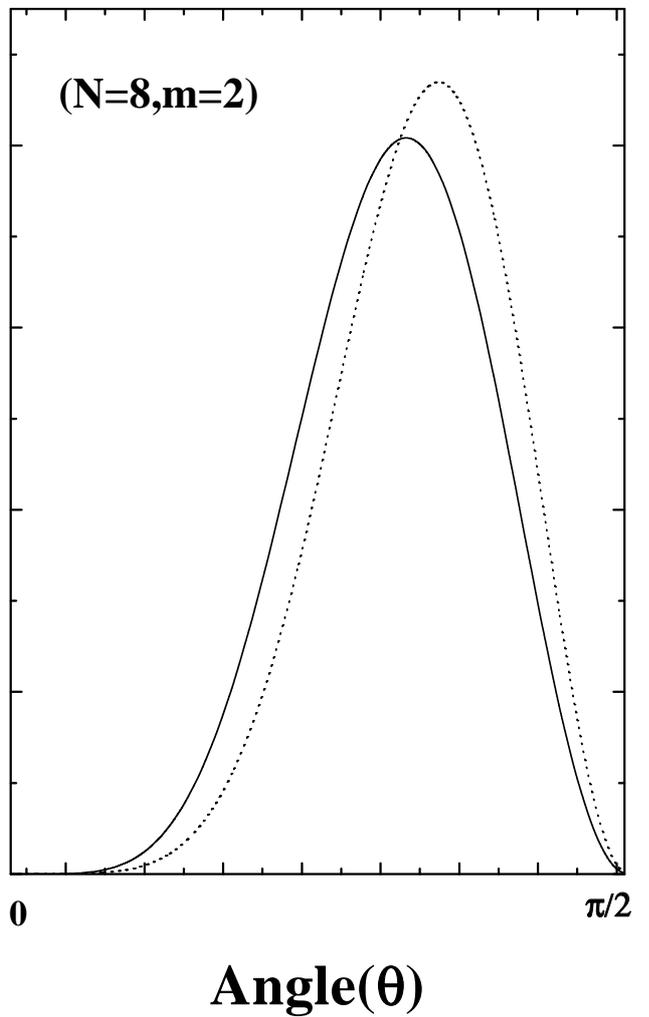
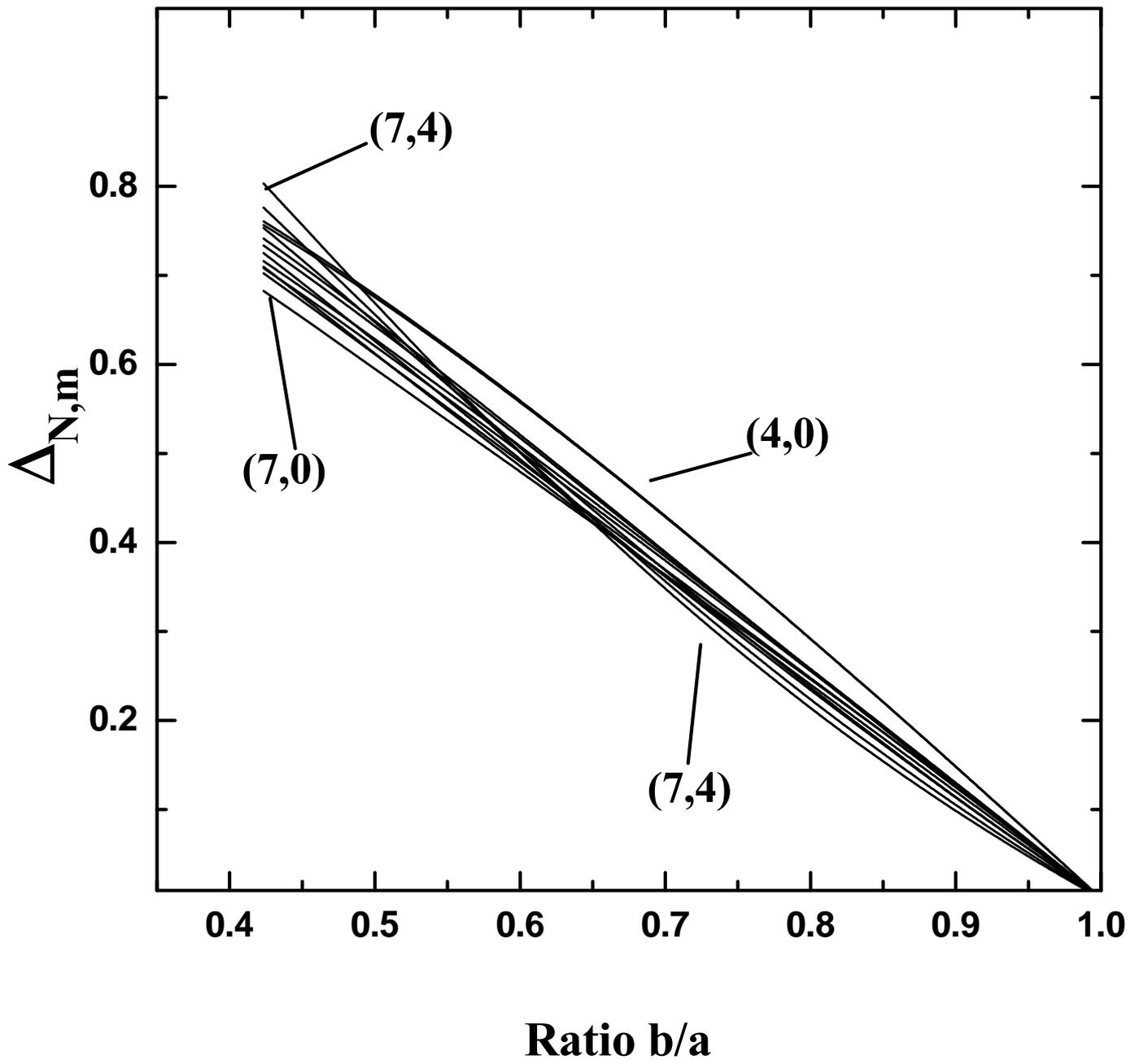
\end{document}